\newif\ifdraft
\draftfalse

\documentclass[a4paper]{llncs}
\usepackage{tikz}
\usepackage{listings}
\lstMakeShortInline[basicstyle=\tt]|
\usepackage{local}

\ifdraft
\else
\usepackage{butterma}
\idline{The final publication of this paper is available at www.springerlink.com}
\makeatletter
\def\ps@first{\let\@mkboth\@gobbletwo
     \def\@oddfoot{\reset@font\scriptsize
     \vbox to\z@{\parindent=\z@\vss
     \@idline \unskip, foo\\
     }}\let\@evenfoot\@oddfoot}
\makeatother
\fi

\usepackage[T1]{fontenc}
\usepackage[utf8]{inputenc}
\usepackage{courier}

\ifdraft
\usepackage[today,eso-foot]{svninfo}
\usepackage[show]{ed}
\else
\usepackage[hide]{ed}
\fi

\usepackage{amstext,amsfonts,amssymb}
\usepackage{paralist}
\usepackage{url}
\usepackage{wrapfig}
\usepackage{stex-logo}
\usepackage{xspace}

\usepackage[pdfpagescrop={92 112 523 778},a4paper=false]{hyperref}

\newcommand{\sd}{{\small{SAMSDocs}}\xspace}

\newcommand{\vmodel}{V-Model\xspace}
\newcommand{\latexml}{{\LaTeX}ML}
\newif\iflong
\longtrue

\def\name#1{{\textsc{#1}}\index{#1@{\sc #1}}}
\def\myCitation#1{{\small ``{\emph{#1}}''}}

\def\schutzfeld{safety zone\xspace}
\def\sams{SAMS\xspace}

\def\omdoc{OMDoc\xspace}

\newcommand{\stexSD}{\stex-SD\xspace}
\newcommand{\explicate}{formalize}

\title{Dimensions of Formality:\\
  A Case Study for MKM in Software Engineering}%
\author{Andrea Kohlhase\inst{1} and Michael Kohlhase\inst{2} and Christoph Lange\inst{2}}
\institute{German Research Center for Artificial Intelligence (DFKI)\\
  \email{Andrea.Kohlhase@dfki.de}
\and
Computer Science, Jacobs University Bremen\\
\email{\{m.kohlhase,ch.lange\}@jacobs-university.de}}

\begin{document}
\ifdraft
\svnInfo $Id: multiform.tex 1538 2010-04-28 12:33:23Z clange $
\svnKeyword $HeadURL: https://svn.kwarc.info/repos/swim/doc/metadata/multiform.tex $
\fi
\maketitle

\begin{abstract} 
  We study the formalization of a collection of documents created for a Software
  Engineering project from an MKM perspective. We analyze how document and collection
  markup formats can cope with an open-ended, multi-dimensional space of primary and
  secondary classifications and relationships. We show that RDFa-based extensions of MKM
  formats, employing flexible ``metadata'' relationships referencing specific
  vo\-ca\-bu\-la\-ries for distinct dimensions, are well-suited to encode this and to put
  it into service. This {\explicate}d knowledge can be used for enriching interactive
  document browsing, for enabling multi-dimensional metadata queries over documents and
  collections, and for exporting Linked Data to the Semantic Web and thus enabling further
  reuse.
\end{abstract}

\section{Introduction}\label{sec:intro}
The field of Mathematical Knowledge Management (MKM) tries to model mathematical objects
and their relationships, their creation and publication processes, and their management
requirements. In~\cite[237 ff.]{CF:ReviewMKM09} \name{Carette} and \name{Farmer}
analyzed {\myCitation{six major lenses through which researchers view MKM}}: the document,
library, formal, digital, interactive, and the process lens. Quite obviously, there is a
gap between the formal aspects \{``library'', ``formal'', ``digital''\} -- related to
machine use of mathematical knowledge -- and the informal ones \{``document'',
``interactive'', ``process''\} -- related to human use.

\begin{wrapfigure}{r}{.48\textwidth}\vspace*{-2em}
  \includegraphics[width=.48\textwidth]{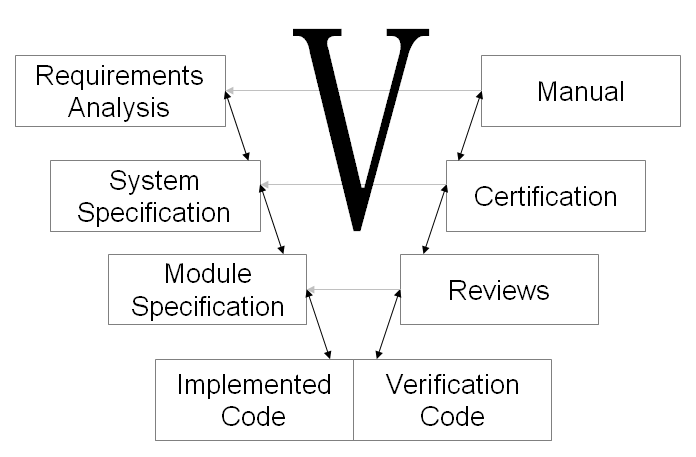}\vspace*{-1em}
  \caption{Documents in the V-Model}\label{fig:V-model}\vspace*{-2.4em}
\end{wrapfigure}

In the FormalSafe project~\cite{URL:FormalSafe} at the German Research Center for
Artificial Intelligence (DFKI) Bremen a main goal is the integration of project documents
into a computer-supported software development process. MKM techniques are used to bridge
the gap between informally stated user requirements and formal verification. One of the
FormalSafe case studies is based on the documents of the {\sams} project
(``Sicherungskomponente f\"ur Autonome Mobile Systeme [Safety Component for Autonomous
Mobile Systems]'', see~\cite{sams:SafeCert08}) at DFKI. The {\sams} objective was to
develop a safety component for autonomous mobile service robots and to get it certified as
SIL-3 standard compliant in the course of three years. On the one hand, certification
required the verification of certain safety properties in the code documents with the
proof checker Isabelle~\cite{Nipkow-Paulson-Wenzel:2002}. On the other hand, it
necessitated the software development process to follow the \vmodel
(fig.~\ref{fig:V-model}). This mandates e.\,g.\ that relevant document fragments get
justified and linked to corresponding fragments in a successive document refinement
process (the arms of the `V' from the upper left over the bottom to the upper right and
between arms in fig.~\ref{fig:V-model}).

The collection of {\sams} documents (we call it ``{\bf{\sd}}'' \cite{SAMSDocs}) promised
an interesting case study for FormalSafe as system development with respect to the
{\vmodel} regime resulted in a highly interconnected collection of design documents,
certification documents, code, formal specifications, and formal proofs. Furthermore, it
was supposed that adding semantics to {\sd} would be comparatively easy as it was
developed under a strong formalization pressure.

In this paper we report on --- and draw conclusions from --- the {\sd} formalization,
particularly the formalization of its {\LaTeX} documents. In section
~\ref{sec:formalization}, we document the process and detect inherent, distinct formality
levels and the multi-dimensionality of the {\explicate}d structures. Real information
needs (drawn from three use cases in the \sams context) turn out in
section~\ref{sec:usecases} to be multi-dimensional. This motivates our exploration of
multi-dimensional markup in section~\ref{sec:markup}. Section~\ref{sec:semanticweb}
showcases the feasibility of multi-dimensional services with MKM technology enabled by
multi-dimensional structured representations and section~\ref{sec:conclusion} concludes
the paper.

\section{Dimensions of Formality in SAMSDocs}\label{sec:formalization}
In this paper, we are especially interested in the question
{\bf\emph{``What should we sensibly {\explicate} in a document collection and can MKM methods cope?''}}. Note that
we understand ``to {\explicate}'' as ``making implicit knowledge explicit'' and not as
``to make s.th. fully formal''.

The {\sams} project was organized as a typical Software Engineering project, its collection
of documents {\sd} therefore has a prototypical composition
\begin{wrapfigure}l{3.8cm}\vspace*{-2em}
\footnotesize
\lstDeleteShortInline|
\begin{tabular}[t]{|l|l|l|}\hline
{\bf{Format}}        & {\bf{Files}}     &  {\bf{\#}} \\\hline
\LaTeX               & \texttt{*.tex}   & 251        \\\hline
MS Word              & \texttt{*.doc}   & 61         \\\hline
Isabelle             & \texttt{*.thy}   & 33         \\\hline
Misra-C Code         & \texttt{*.c}     & 40         \\\hline
\end{tabular}\vspace*{-.5em}
\lstMakeShortInline[basicstyle=\tt]|
\caption{\sd}\label{fig:samsdocs}\vspace*{-2.4em}
\end{wrapfigure}
 of distinct document
types like contract, code, or manual. Thus, {\sd} presents a good base for a case study
with respect to our question. In fig.~\ref{fig:samsdocs} we can see the concrete distribution over used document formats in {\sd}. Requirements analysis, system
and module specifications, reviews, and the final manual were mainly written in {\LaTeX},
only roughly a sixth in MS Word. The implementation in Misra-C contains Isabelle theorem
prover calls.  

The first stinging, but unsurprising observation was that the {\bf\emph{level of
    formality}} of the documents in {\sd} varies considerably --- because distinct purposes
create distinct formality requirements. For instance, the contract document serves as
communication medium between the customer and the contractor. Here, underspecification is
an important tool, whereas it is regarded harmful in the fine-granular module
specifications and a fatal flaw in input logic for a theorem prover. Since this issue was
already present in the set of {\LaTeX} documents, we focused on just these.

For the formalization of this subset in {\sd} we used the \stex
system~\cite{Kohlhase:ulsmf08}, a semantic extension of {\LaTeX}. It offers to both
publish documents as high-quality human-readable PDF {\emph{and}} as formal
machine-processable \omdoc~\cite{Kohlhase:OMDoc1.2} via {\latexml}~\cite{StaKoh:tlcspx10}.
Our formalization process revealed early on that previous {\stex} applications (based on
{\omdoc}~1.2) were too rigid for a stepwise semantic markup. But fortunately, {\stex} also
allows for the {\omdoc}~1.3 scheme of metadata via RDFa~\cite{AdidaEtAl08:RDFa}
annotations (see~\cite{Kohlhase:OMDoc1.3}). In particular, we could `invent' our own
vocabulary for markup on demand without extending OMDoc.  This new vocabulary consists of
{\sd}-specific metadata properties and relationship types.  We call the process of adding
this {\emph{pre}}-formal markup to {\sd} {\textbf{(semantic) preloading}}. Concretely, we
extended \stex to {\stexSD} (\stex for \sd) by adding {\latexml} bindings for all \sams
specific {\TeX} macros and environments used in \sd, thus enabling the conservation of the
original PDF document layouts at the same time as the generation of meaningful OMDoc.

\begin{figure}[h]
  \includegraphics[width=\textwidth]{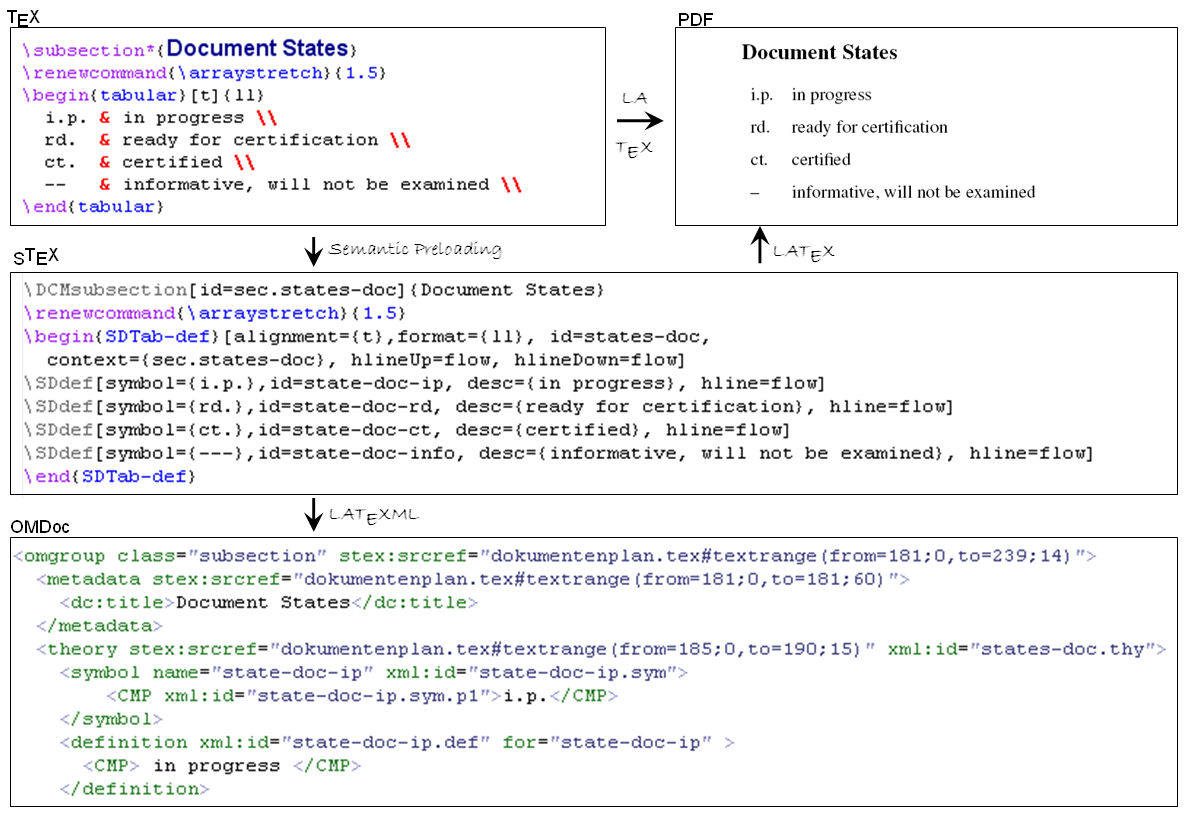}
  \caption{The Formalization Workflow with \protect\stexSD $\lbrack$ translated by the
    authors $\rbrack$}\label{fig:semantification13}
\end{figure}
Let us look at an example for such an {\stex} extension within our formalization workflow
(see fig.~{\ref{fig:semantification13}}). We started out with a {\TeX} document (upper
left), which compiled to the PDF seen on the upper right. Here, we have a simple,
two-dimensional table, which is realized with a {\LaTeX} environment
|tabular|. Semantically, this table contains a list of symbols for document states with
their definitions, e.\,g.\ ``i.\,B.''  for ``in Bearbeitung [in progress]''. As such definition
tables were used throughout the project, we developed the environment |SDTab-def| and the
macro |SDdef| as {\stex} extensions. We determined the {\omdoc} output for these to be a
symbol together with its definition element (for each use of |SDdef| in place of the
resp. table row) and moreover, to group all of them into a theory (via using
|SDTab-def|). Preloading the {\TeX} table by employing |SDTab-def| and |SDdef| turned it
into an {\stex} document (middle of fig.~{\ref{fig:semantification13}}) while keeping the
original PDF table structure. Using {\latexml} on this {\stex} document produces the
{\omdoc} output shown in the lower area of fig.~{\ref{fig:semantification13}}.

Mathematical, structural relationships have a privileged state in {\stex}: their command
sequence/environment syntax is analogous to the native XML element and attribute names in
{\omdoc}. Since many objects and relationships induce formal representations for Isabelle,
it seemed possible to semantically mark them up with a logic-inspired structure. But in
the formalization process it soon became apparent that (important) knowledge implicit in
{\sd} did not refer to the `primary' structure aimed at with the use of {\stex}. Instead,
this knowledge was concerned with a space of less formal, `secondary' classifications and
relationships. Thus, our second observation pertains to the substance of
formalizations. Even though we wanted to find out {\emph{what}} we can sensibly formalize,
we had assumed this to mean {\emph{how much}} we can sensibly formalize. Therefore, we
were rather surprised to find distinct {\bf\emph{formality structures}} realized in our
{\stex} extension. In the following we want to report on these structures.

We grouped the macros and environments of {\stexSD} in fig.~\ref{fig:STEX-SD} according to what induced
them. Particularly, we distinguished the following triggers:
\begin{compactitem}\item ``{\emph{object}}s'' --- document fragments viewed as
autonomous elements --- and 
\item their net of relationships via the {\emph{collection}},
\item {\emph{document}}s and
\item their {\emph{organization}}al handling, and
\item the {\emph{project}} itself and thus, its own scheme of meaningful
  relationships. \end{compactitem}

\begin{wrapfigure}{l}{2.5cm}\vspace*{-2ex}
\fbox{\includegraphics[width=2.5cm]{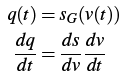}}\vspace*{-2ex}
\caption{$s$ is Bra\-king Distance?}\label{fig:kettenregel}\vspace*{-5ex}
\end{wrapfigure}

\noindent For instance,  in the system specification we marked a recap of a definition of the braking
distance function for straight-ahead driving $s_G$ as an object and referenced it from
within the assertion seen in fig.~\ref{fig:kettenregel}. In the  module specification
$s_G$ was then meticulously specified. This document fragment is connected to the original
one via a refinement-relationship from the \vmodel, which determined the creation process
of the collection. Documents induce layout structures like sections or subsections and
they are themselves organized for example under a version management scheme. In the
workflow in fig.~\ref{fig:semantification13} we already showcased a project-specific
element, the definition table, with its meaning. Interestingly, we cannot compare
formality in one group with the formality in another. For example, we cannot decide
whether a document completely marked up with the object-induced structures is more formal
than one fully semantically enhanced by the version management markup. As these grouped
structures only interact relatively lightly, we can consider them as independent
{\bf{dimensions of a formality space}} that is reified in the formalization process of a
document collection.

Concretely, {\stexSD} covers the following dimensions and consists of the listed extension
macros/environments (with attributes in $\lbrack\cdot\rbrack$ where sensible):
\begin{figure}[h]
  \includegraphics[width=\textwidth]{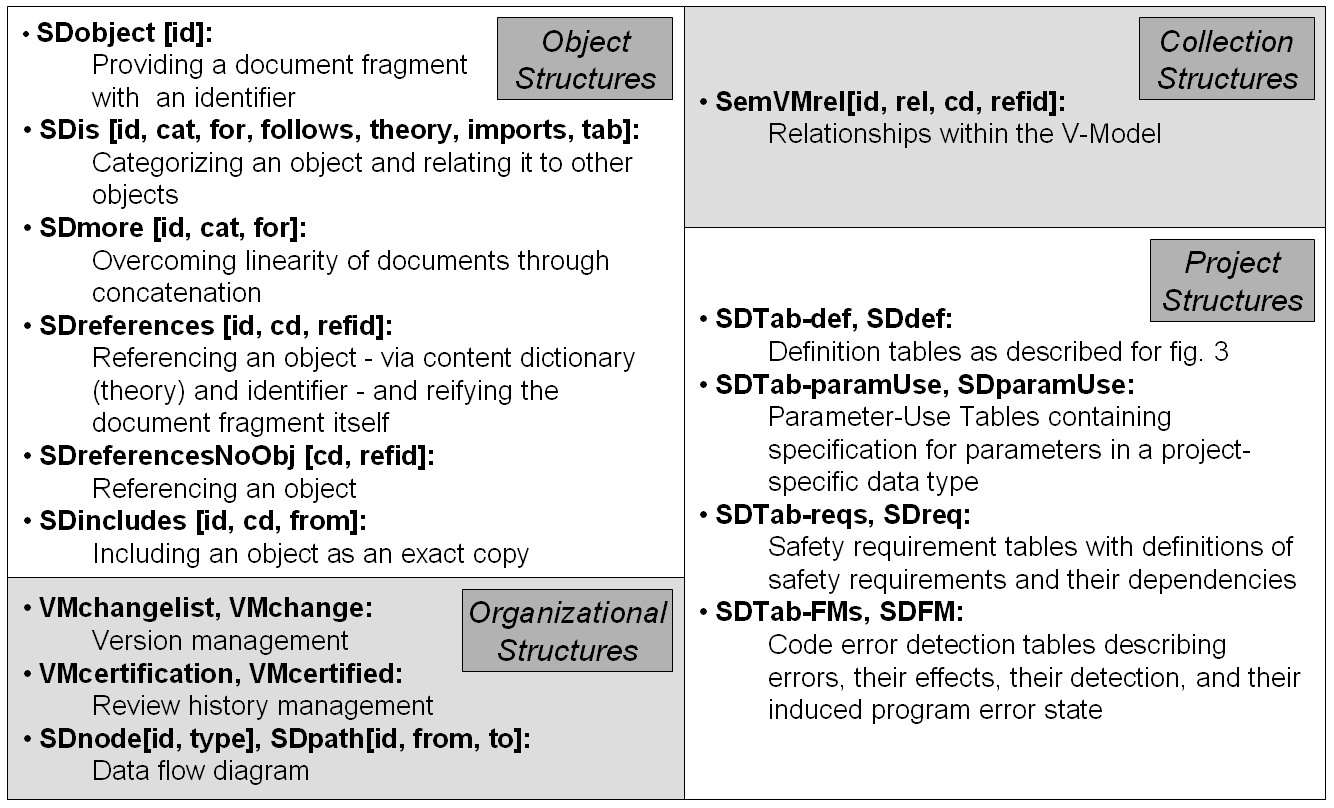}
  \caption{Formality Dimensions in \protect\stexSD}\label{fig:STEX-SD}
\end{figure}

Formalizing object structures is not always obvious, since many of the documents contain
recaps or previews of material that is introduced in other documents/parts (e.\,g.\ to
make them self-contained). Compare for example fig.~\ref{fig:kettenregel}
\begin{wrapfigure}{r}{3.5cm}\vspace*{-5ex}
\fbox{\includegraphics[width=3.4cm]{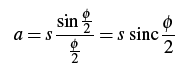}}\vspace*{-2ex}
\caption{Yet another Braking Distance $s$?}\label{fig:sehnensatz}\vspace*{-5ex}
\end{wrapfigure}
 and fig.~\ref{fig:sehnensatz}, which are
actually clippings from the system specification ``\texttt{Konzept\-Bremsmodell.pdf}''.
Note the use of $s$ resp. $s_G$, {\emph{both}} pointing in fig.~\ref{fig:kettenregel} to
the braking distance function for straight-ahead driving (which is obvious from the local
context), whereas in fig.~\ref{fig:sehnensatz} $s$ represents the general arc length
function of a circle, which is different in principle from the braking distance, but
coincides here.

We also realized that {\stex} itself had already integrated another formality dimension
besides the logic-inspired one, the one concerned with document layout:
\begin{figure}[h]
  \includegraphics[width=\textwidth]{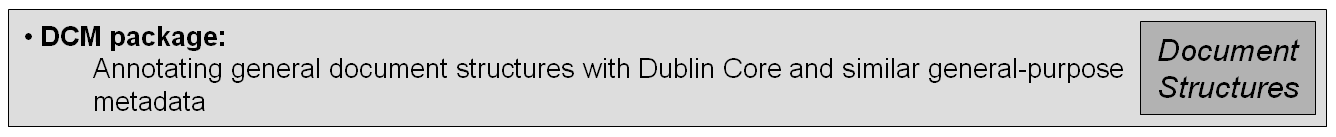}
  \caption{The Document Formality Dimension in \protect\stex}\label{fig:STEX-DCM}
\end{figure}
A typical document layout is structured into established parts like sections or
modules. If we want to keep this grouping information in the formal XML document, we might
use {\stex}'s {\texttt{DCM}} package for annotating general document structures with
Dublin Core (cf.~\cite{DCMI:dcmes08:tr}) and similar general-purpose metadata. In the
{\stex} box in fig.~\ref{fig:semantification13} we find for example the command
{\texttt{DCMsubsection}} with attributes containing the title of the subsection and an
identifier that can be used in the usual {\LaTeX} referencing scheme.


\medskip Finally, we would like to remark that the \stexSD preloading process was executed
as {\myCitation{in-place formalization}}~\cite{ShipmanMcCall:IncrementalFormalization}. It
frequently considered several of the above dimensions for the object at hand at the same
time. Therefore, the often applied metaphor of ``formalization steps'' does not mirror the
formalization process in our case study. We found that the important aspect of the
formalization was not its sequence per se, which we consider particular to the \sd
collection, but the fact that the metaphor of `steps' only worked within each single
dimension of formality. In particular, there is no scale for formalization progress as
distinct formality levels in distinct formality dimensions existed in a document at one
point in time.
 
\section{Multi-Dimensional Information Needs}\label{sec:usecases}
We have shown that the formalization of knowledge results in an open-ended,
multi-dimensional space of primary and secondary classifications and relationships. But
are multi-dimensional document formalizations beneficial for services supporting real
users?  Concretely, we envision potential questions in the {\sams} context and 
services that retrieve and display answers based on the multi-dimensional markup of {\sd}.

Let us first take a {\bf{programmer}}'s perspective. Her main information source for the
programming task will stem from the module specification. But while studying it the following questions
might arise:
\begin{compactenum}[(i)]  
\setcounter{enumi}{0}
\item\label{item:lookupDefinition} What is the definition for a certain (mathematical)
  symbol?\footnote{See fig.~\ref{fig:kettenregel} and~\ref{fig:sehnensatz} for two
    symbols having the same appearance but different meanings.}
\item\label{item:alreadyImplemented} How much of this specification has already been implemented?
\item\label{item:proofAlreadyVerified} In what state is the proof of a specific equation,
  has it already been formally verified so that it is safe to ground my implementation on
  it?
\item\label{item:callForHelp} Whom can I ask for further details?
\end{compactenum} 

Assuming multi-dimensional markup an information retrieval system can supply
useful responses. For example, it can answer~(\ref{item:lookupDefinition}) if technical
terms in natural language are linked to the respective formal mathematical symbols they
represent. For replies to~(\ref{item:alreadyImplemented})
and~(\ref{item:proofAlreadyVerified}) we note that, if all collection links are merged
into a graph, their original placement and direction no longer makes a
difference. So if we have links from the Isabelle formalization to the respective C code
and links from this C code to a specification fragment, as realized in the \vmodel
structure of \sd, we can follow the graph from the specification through to the state of
the according proof. Drawing on the \vmodel links combined with the semantic version
management or the review logs, the system can deduce the answer
for~(\ref{item:callForHelp}): The code in question connects to a specification document
that has authors and reviewers. This service can be as fine-grained as one is willing to
formalize the granularity of the version and review management. If we admit further
dimensions of markup into the picture, then the system might even find persons with similar interests
(e.\,g.\ expressed in terms of the FOAF vocabulary), as has been investigated in detail
for expert finder systems~\cite{SWJL:ExpertLinkedData10}.

Now, we take a more global perspective, the one of a {\bf{project manager}}. She might be
concerned with the following issues:
\begin{compactenum}[(i)]
  \setcounter{enumi}{4}
\item\label{item:SEprocess}{\emph{Software Engineering Process}:} How much code has been
  implemented to satisfy a particular requirement from the contract?  Has the formal code
  structure passed a certain static analysis and verification?  She does not want to
  inspect that manually by running Isabelle, thus, she needs high-level figures of,
  e.\,g., the number of mathematical statements without a formally verified proof.
\item\label{item:certification}{\emph{Certification}:} What parts of the specification,
  e.\,g.\ requirements, have changed since the last certification?  What other parts does
  that affect, and thus, what subset of the whole specification has to be re-certified?
\item\label{item:humanCapital}{\emph{Human Capital}:} Who is in charge of a document?  How
  could an author be replaced if necessary, taking into account colleagues working on the
  same or on related documents – such as previous revisions of the same document, or its
  predecessor in terms of the \vmodel, i.\,e.\ the document that is refined by the current
  one?
\end{compactenum}
Exploiting the multi-dimensionality of formalized knowledge, it becomes obvious how the issues can be tackled.  

Finally, we envision a {\bf{certifier}}'s information needs.  For inspection, she might
first be interested in getting an overview, such as a list of all relevant concepts in the
contract document.  Then, she likes to follow the links to the detailed specification
and further on to the actual implementation.  For more information, she likes to contact the
project investigator instead of the particular author of a code snippet. The certifier
also needs to understand what parts of the whole specification are subject to a requested
re-certification. Her rejection of a certain part of a document also affects all
elements in the collection that depend on it. Again, a system can easily support a
certifier's efficiency by combining the {\explicate}d information of distinct formality
dimensions.

These use scenarios in a Software Engineering project clearly show that multi-dimensional
markup is useful, since multi-dimensional queries serve natural information needs. To
answer such queries, we need to represent multi-dimensional information in MKM formats. 

\section{Multi-Dimensional Markup}\label{sec:markup}

Structured representations are usually realized as files marked up in formats that reflect
the primary formalization intent and markup preferences of the formalizer. In the
evaluation of document formats it is thus important to realize that every representation
language concentrates on only a subset of possible relationships, which it treats with
specific language constructs. Note that therefore the formality space of a semantically
enhanced document is very often reduced to this primary dimension. On the formal side, for
example, a plethora of system-specific logics exist.  Furthermore, formal systems
increasingly contain custom modularization infrastructures, ranging from simple facilities
for inputting external files to elaborate multi-logic theory
graphs~\cite{MosMaeLue:thts07}. Collections of informal documents, on the other side, are
often structured by application-specific metadata like the Math Subject
Classification~\cite{AMS:MSC2010} or the {\vmodel} relations as in our case study.

No given format can natively capture \emph{all} aspects of the domain via special-purpose
markup primitives. It has to relegate some of them to other mechanisms like the {\stexSD}
extension for the formalization of {\sd}, if more dimensions
of the formality space than the primary one are to be covered. In representation formats
that support fragment identifiers --- e.\,g.\ XML-based ones --- these relationships can be
expressed as stand-off markup in RDF (Resource Description Framework~\cite{w3c:rdf}), i.\,e.,
as subject-predicate-object triples, where subject and object are URI references to a
fragment and the predicate is a reference to a relationship specified in an external
vocabulary or ontology\footnote{The difference between ``vocabulary'' and ``ontology'' is
  not sharply defined.  Vocabularies are often developed in a bottom-up community effort
  and tend to have a low degree of formality, whereas ontologies are often designed by a
  central group of experts and have a higher degree of formality.  Here, we use
  ``vocabulary'' in its general sense of a set of terms from a particular domain of
  interest.  This subsumes the term ``ontology'', which we will reserve for cases that
  require a more formal domain model.}. As we have XML-based formats for informal
documents (e.\,g.\ XHTML+MathML+SVG) and formal specifications (OpenMath or Content MathML),
we can in principle already encode multi-dimensional structured representations, if we
only supply according metadata vocabularies for their structural relationships. Indeed
this is the basic architecture of the ``Semantic Web approach'' to eScience, and much of
the work of MKM can be seen as attempts to come up with good metadata vocabularies for the
mathematical/scientific domain.

Since RDF stand-off markup is notoriously difficult to keep up to date, {\bf{RDFa}}
\cite{AdidaEtAl08:RDFa} has been developed: A set of attributes for embedding RDF
annotations into XML-based languages, originally XHTML.  On the one hand, RDFa serves as
an enabling technology for making XML-based languages extensible by inter- and
intra-document relationships.  On the other hand, RDFa serves as a vehicle for document
format interoperability. All relationships from a format $F$ that cannot be natively
represented in a format $F^{\prime}$ can be represented as RDFa triples, where the
predicate is from an appropriately designed metadata vocabulary that describes the format
$F$. For instance, an {\omdoc} |<theory>| element can be represented as |<div typeof="http://omdoc.org/ontology#Theory">| in XHTML, using the OMDoc
ontology~\cite{OMDocDocOnto:web}. Support of RDFa relationships make all XML-based formats
theoretically equivalent, if they allow fine-grained text structuring with elements like
XHTML's \texttt{div} or \texttt{span} everywhere (so that arbitrary text fragments can be
turned into objects). In particular, they become {\bf{formats for multi-dimensional markup}} as
respective other dimensions can always be added via RDFa. We have detailed the necessary
extensions for the {\omdoc} format in~\cite{Kohlhase:OMDoc1.3}, so that analogous
extensions for any of the XML-based formats used in the MKM community should be rather
simple to create.

Note that the pragmatic restriction to XML-based representation formats is not a loss of
generality. In the MKM sphere the three classes of non-XML languages used are
computational logics, {\TeX}/\LaTeX, and PostScript/PDF.  We see computational logics as
compact front-end formats that are optimized for manual input of formal structured
representations; it is our experience that these can be transformed into the XML-based
OpenMath, MathML, or {\omdoc} without loss of information (but with a severe loss of
notational conciseness). We consider {\TeX}/{\LaTeX} as analogous for informal structured
representations; they can be transformed to XHTML+MathML by the {\latexml} system. The
last category of formats are presentation/print-oriented output and archival formats where
the situation is more problematic: PostScript (PS) is largely superseded by PDF which
allows standard document-level RDF annotations via XMP and the finer-granular annotations
we need for structured representations via extensions as in~\cite{Groza:SALT-claims07}
or~\cite{Eriksson:SemanticDocument2007}. But PS/PDF are usually generated from other
formats (mostly office formats or {\LaTeX}), so that alternative generation into XML-based
formats like XHTML or {\omdoc} can be used.

Note as well that a dimension typically corresponds to a vocabulary.  In the course of the
\sd case study, most vocabularies have initially been implemented from scratch in a
project-specific ad hoc way. But they can be elaborated towards ontologies via {\stex} and
these can be translated to RDF-based formats that automated reasoners
understand~\cite{KohKohLan:ssffld10}. An alternative is reusing existing ontologies. This
has the advantage that they are more widely used and thus, reusable services may already
have been implemented for them. For instance, there already exists a vocabulary that
defines basic properties of persons and organizations: FOAF (Friend of a
Friend~\cite{FOAF:spec}).  The widely known Dublin Core element set is also available as
an ontology~\cite{DCMI:dcmes08:tr}. DCMI Terms~\cite{DCMI:dcmi-terms:tr}, a modernized and
extended version of the Dublin Core element set, offers a basic vocabulary for revision
histories – but not for reviewing and certification. DOAP (Description of a
Project~\cite{DOAP:web}) describes software projects, albeit focusing on the top-level
structure of public open source projects. {\name{Lin}} et al.\ have developed an ontology
for the requirements-related parts of the \vmodel (cf.\ \cite{LFB:ReqOntoEngDesign96}).
{\name{Happel}} and {\name{Seedorf}} briefly review further ontologies about Software
Engineering~\cite{HS:OntologiesSWEng}.  As, e.\,g.\ the \sd vocabularies can be integrated
with existing ontologies by declaring appropriate subclass or equivalence relationships,
services can make use of the best of both worlds.

\section{Multi-Dimensional Services with MKM Technology}\label{sec:semanticweb}

We will now study an avenue towards a concrete implementation of services based on the use
cases described in sect.~\ref{sec:usecases} to show how MKM technologies can cope with
multi-dimensional information needs demonstrating their feasibility.  Concretely, we will
study the task of project manager Nora to find a substitute for employee Alice.  All
required information is contained in the {\stexSD} documents. To abstract from the
particulars of \stex/\omdoc RDFa encoding --- e.\,g.\ the somewhat arbitrarily chosen
direction of the relations or the interaction of metadata relations with the document and
the special markup for the mathematical dimension --- we extract a uniform RDF
representation of the embedded structures, which can then be queried in the SPARQL
language~\cite{PruSea08:sparql}. Listing~\ref{lst:substitute} shows the necessary query in
all detail.

\begin{lstlisting}[language=sparql,caption={Finding a Substitute for an Employee via the \vmodel},label={lst:substitute},escapeinside={[]},columns=fixed]
# declaration of vocabulary (= dimension) namespace URIs
PREFIX vm:    <http://www.sams-projekt.de/ontologies/VersionManagement[\#]>
PREFIX omdoc: <http://omdoc.org/ontology[\#]>                                 # OMDoc
PREFIX semVM: <http://www.sams-projekt.de/ontologies/V-model[\#]>
PREFIX dc:    <http://purl.org/dc/elements/1.1/>                     # Dublin Core
PREFIX xsd:   <http://www.w3.org/2001/XMLSchema[\#]>           # XML Schema datatypes

SELECT ?potentialSubstituteName WHERE {
  # for each document Alice is responsible for, get all of its parts
  # i.e. [\tt any] kind of semantic (sub)object in the document
  ?document vm:responsible <.../employees[\#]Alice> ;
            omdoc:hasPart  ?object .

  # find other objects that are related to each [\tt ?object]
  # 1. in that [\tt ?object] refines them via the V-model
  { ?object semVM:refines ?relatedObject }
  UNION
  # 2. or in that they are other mathematical symbols defined in terms
  #      of [\tt ?object] (only applies if [\tt ?object] itself is a symbol)
  { ?object omdoc:occursInDefinitionOf ?relatedObject }

  # find the document that contains the related object and the person 
  # responsible for that document ...
  ?otherDocument omdoc:hasPart  ?relatedObject ;
                 dc:date        ?date ;
                 vm:responsible ?potentialSubstitute .
  # (only considering documents that are sufficiently up to date)
  FILTER (?otherDocument > "2009-01-01"^^xsd:date)

  # ... and the real name of that person
  ?potentialSubstitute foaf:name ?potentialSubstituteName .
}
\end{lstlisting}

In this query we assume that Alice's FOAF profile is a part of our collection, having the
URI \url{.../employees#Alice}.  Nora retrieves all documents in the collection for which
Alice is known to be the responsible person.  For any object $O$ in each of these documents
(e.\,g.\ the detailed specification of the braking distance function for straight-ahead
driving $s_G$ from fig.~\ref{fig:kettenregel}), she selects those objects that
are refined by $O$ in terms of the \vmodel (e.\,g.\ the general braking distance $s$).
Additionally, she considers the mathematical dimension and selects all objects that are
related to $O$ by mathematical definition, e.\,g.\ the braking function that uses $s_G$.  Of any such related object, Nora finds out to what document it belongs.
She is only interested in recent documents and therefore filters them by date.  Finally,
she determines the responsible persons via the version management links, and gets
their names from their FOAF descriptions.  The assumption behind this query is that, if,
for example, Pierre is responsible for the specification that introduces the general
braking distance $s$, which Alice has refined, Pierre can be considered as a substitute
for Alice.  Note that getting the answer draws on the collection structures of {\sd}
(\vmodel), on the mathematical structures, as well as on the organizational structures
(version management).  It is easy to imagine how additional formality dimensions can be
employed for increasing precision or recall of the query, or for ranking results.
Consider, for example, another filter that only accepts as substitutes employees who have
never got a document rejected in any previous certification.

The complexity of the query in listing~\ref{lst:substitute} is directly caused
by the complexity of the underlying multi-dimensional structures and the non-triviality of answering
high-level project management queries from the detailed information in {\sd}. As users like Nora would not want to deal with a machine-oriented query language, we have
developed a system that integrates versioned storage of semantic document collections with
human-oriented presentation with embedded interactive
services~\cite{DKLRZ:PubMathLectNotLinkedData10}. Thus, the rendered documents serve as
command centers for executing queries and displaying results\footnote{In particular, the
  rendered XHTML+MathML also preserves the original semantic structure as parallel MathML
  markup and RDFa annotations, so that a suitable browser plugin can dynamically generate
  interaction points for semantic services; see~\cite{KohKohLan:ssffld10} for
  details.}. They provide access to queries in two ways: Queries with a fixed
structure that have to be answered recurringly will be made available right in the
(rendered) documents in appropriate places. This is the case with our employeee substitution query: This month, Alice may be ill, whereas next month, Bob may be on holiday.
Access to this query can be given wherever an employee or a reference to an employee
occurs in a document.  Alternatively, non-prefabricated queries can be composed more
intuitively on demand using a visual input form.

These examples show that
multi-dimensional queries like the ones naturally coming up in Software Engineering
scenarios (sect.~\ref{sec:usecases}) can be answered with existing MKM technology. Moreover, it illustrates that
multi-dimensional markup affords multi-dimensional services. If we interpret our
dimensions as distinct contexts, our services become context-sensitive, as dimensions can
be filtered in and out. For instance, the context menu of certification documents could be
equipped with menu entries for committing an approval or rejection to the server, which
would only be displayed to the certifier.  The server could then trigger further actions,
such as marking the document that contains a rejected object and all dependencies of that
object as rejected, too. In general, the more dimensions are {\explicate}d in a document,
the more context-sensitive services become available.

\section{Conclusion and Further Work}\label{sec:conclusion}
In this paper we have studied the applicability of MKM technologies in Software
Engineering beyond ``Formal Methods'' (based on the concrete {\sd} document collection and
its formalization). The initial hypothesis here is that contract
documents, design specifications, user manuals, and integration reports can be partially
formalized and integrated into a computer-supported software development process. To test
this hypothesis we have studied a collection of documents created for the development of a
{\schutzfeld} computation, the formal verification that the braking trajectory always lies
in the \schutzfeld, and the SIL3 certification of this fact by a public certification
agency. As the project documents contain a wealth of (informal) mathematical content, MKM
formats (in this case our \omdoc format) are well-suited for this task. During the
formalization of the {\LaTeX} part of the collection, we realized that the documents
contain an open-ended, multi-dimensional space of formality that can be used for
supporting projects --- if made explicit. 

We have shown that RDFa-based extensions of MKM formats, employing flexible ``metadata''
relationships referencing specific vocabularies, can be used to encode this formality
space and put it into service. We have pointed out that the ``dimensions'' of this space
can be seen to correspond to different metadata vocabularies. Note that the distinction
between data and metadata blurs here as, for example, the \omdoc data model realized by
native markup in the \omdoc format can also be seen as \omdoc metadata and could equally
be realized by RDFa annotations to some text markup format, where the meaning of the
annotations is given by the \omdoc ontology.  This ``metadata view'' is applicable to all
MKM formats that mark up informal mathematical texts (e.\,g.\ MathDox~\cite{ccb:MDMDonW} and
MathLang~\cite{KWZ:CmtiM}) as long as they {\explicate} their data model in an
ontology. This observation makes decisions about which parts of the formality space to
support with native markup a purely pragmatic choice and opens up new possibilities in the
design of representation formats. It seems plausible that all MKM formats use native
markup for mathematical knowledge structures (we think of them as primary formality
structures for MKM) and differ mostly in the secondary ones they
internalize. XHTML+MathML+RDFa might even serve as a baseline interchange format for MKM
applications\footnote{Indeed, a similar proposal has been made for Semantic
  Wikis~\cite{VoOr06:wif}, which have related concerns but do not primarily involve
  mathematics.}, since it is minimally committed. Note that if the metadata ontologies are
represented in modular formats that admit theory morphisms, then these can be used as
crosswalks between secondary metadata for higher levels of interoperability. We leave its
development to future work.

The {\explicate}d secondary formality structures can be used for enriching interactive
document browsing and for enabling multi-dimensional metadata queries over documents and
collections.  We have shown a set of exemplary multi-dimensional services based on the
RDFa-encoded metadata, mostly centered around Linked Data approaches based on RDF-based
queries. More services can be obtained by exporting Linked Data to the Semantic Web or a
company intranet and thus enabling further reuse. In particular, the multi-dimensionality
observed in this paper and its realization via flexible metadata regimes in representation
formats allows the knowledge engineers to tailor the level of formality to the intended
applications.

In our case study, the metadata vocabularies ranged from project-specific ones that had to
be developed (e.\,g.\ definition tables) to general ones like the \vmodel vocabulary, for which
external ontologies could be reused later on. We expect that such a range is generally the case
for Software Engineering projects, and that the project-specific vocabularies may stabilize
and be standardized in communities and companies, lowering the formalization effort
entailed by each individual project. In fact we anticipate that such metadata vocabularies
and the software development support services will become part of the strategic knowledge
of technical organizations.

In~\cite[241]{CF:ReviewMKM09} \name{Carette} and \name{Farmer} challenge MKM researchers
by assessing some of their technologies: {\myCitation{A lack of requirements analysis very
    often leads to interesting solutions to problems which did not need solving}}.  With
this paper we hope to have shown that MKM technologies can be extended to cope with ``real
world concerns'' (in Software Engineering). Indeed, industry is becoming more and more
aware of and interested in Linked Data (see e.\,g.~\cite{Servant:LinkingEnterpriseData08}
and~\cite[Question 14]{LinkedDataFAQ:URL}), which boosts relevance to the
multi-dimensional knowledge management methods presented in this paper.

\bibliographystyle{alphahurl}
\bibliography{\macpathkbibs{kwarc},\macpathabibs{ako}}
\end{document}